\documentstyle[12pt]{article}
\begin{document}
\begin{flushright}
WM-97-107\\
DRAFT\\
\today
\end{flushright}
\Large
\centerline
{\bf Millimeter-wave Signature of Strange Matter Stars}
\normalsize
\vskip 2cm
\bigskip
\centerline{John J. Broderick${}^1$, Eugene T. Herrin${}^2$, Timothy P. Krisher
${}^3$, David L.
Morgan${}^4$,}\centerline{ Doris C. Rosenbaum${}^5$, Marc
Sher${}^4$ and Vigdor L. Teplitz${}^5$}
\vskip 1cm
\bigskip
\centerline{${}^1$\it Physics Department, Virginia Polytechnic Institute and
State University, Blacksburg VA 24061}
\centerline{${}^2$\it Geology Department, Southern Methodist University, Dallas
TX, 75275}
\centerline{${}^3$\it Jet Propulsion Lab, California Institute of
Technology, Pasadena CA  91125}
\centerline{${}^4$\it Physics Department, College of William and Mary,
Williamsburg VA  23187}
\centerline{${}^5$\it Physics Department, Southern Methodist University, Dallas
TX, 75275}

\vspace{2cm}

{\narrower\narrower
One of the most important questions in the study of compact objects is the 
nature
of pulsars, including whether they consist of neutron matter or strange quark
matter (SQM).  However, few mechanisms for distinguishing between these 
two possibilities have been proposed.  The purpose of this paper is to show 
that a strange star (one made of SQM) will have a vibratory mode with an
 oscillation frequency of approximately 250 GHz (millimeter wave).  This mode 
corresponds to motion of the center of the expected crust of normal matter 
relative to the center of the strange quark core, without distortion of either.
  Radiation 
from currents generated in the crust at the mode frequency would be a SQM 
signature.  We also consider effects of stellar rotation, estimate power 
emission and signal-to-noise ratio, and discuss briefly possible mechanisms for
 exciting the mode.}

\newpage\section{Introduction} Witten (1984) pointed out 
that strange quark matter (SQM) composed roughly of equal numbers of up, down,
 and strange quarks is more likely to be stable than non-strange quark matter 
(which would have only up and down quarks but is known not to be stable). This
 is because conversion to strange quarks (for $m_s < m_N/3$) lowers the Fermi 
energy. This fact was known to others (see, for example, Bodmer (1971), Friedman
 and McLerran (1978)). Witten, however, went on to suggest that nuggets of 
strange quark matter could be produced in phase transitions in the early 
universe or in supernova explosions and gave a possible scenario for the 
former, modified versions of which are still under debate; see for example, 
Cottingham et al. (1994).  Witten raised the possibility that such nuggets 
could solve the cosmological dark matter problem by evading the bound on the 
cosmological baryon density from the abundance of primordial deuterium.  Fahri 
and Jaffe (1984) considered in some detail the properties of such nuggets as a 
function of nucleon number.  de Rujula and Glashow (1984) considered 
terrestrial effects, on land, in the sea and in the air, from incident strange 
quark nuggets. 

Alcock, Fahri and Olinto (1986) discussed in depth the 
conversion of neutron stars to strange stars and the structure and properties 
of the latter.  This work was continued by a number of others and the subject 
was reviewed thoroughly at a 1991 conference (Madsen and Haensel, 1991). The 
consensus is that, if SQM is stable, all ``neutron stars" should in reality be
 strange quark stars. In a recent paper, Kettner et al. (1995) have explored 
the nature of strange quark stars in further detail and computed important 
properties at non-zero temperatures;  this paper takes as its point of 
departure that paper to which we refer as KWWG. 

The object of this paper is to 
speculate on a possible millimeter wave radio signal that might be a signature 
of a strange quark star.  Because of the fact that there are few observable 
differences between classical neutron stars and strange quark stars, such a 
signal could aid in identifying strange quark stars.  As far as we are aware, 
to date only the existence of pulsars with shorter periods than permitted to 
classical neutron stars (Frieman and Olinto,1989; Glendenning, 1989) and 
differences in cooling rates (Benvenuto and Vucetich, 1991) have been 
discussed in detail in the literature.  While our proposal is speculative, the
impact of detecting strange quark matter would be so great that we believe it 
important to raise it for discussion and, hopefully, for observational efforts.
Our proposal addresses the case generally contemplated by 
most workers that the strange quark core is surrounded by a crust
of normal matter on the order of $10^{-5}M_{\odot}$.  We note
that, for the case in which the core is almost nude except for
an atmosphere of normal matter of much smaller mass, Usov (1997)
has recently proposed an X-ray signal.

The plan of the paper is as follows:  in the remainder of this section, we 
summarize the features of the KWWG strange star model relevant to our work.  In
 Section II, we consider the frequency of the vibrational mode in which the 
crust of hadronic matter vibrates as a single entity, without distortion, with 
respect to the strange quark core.  Radio waves generated from this vibration 
constitute our proposed signal.  In Section III, the effects of the rotation of
 the strange star are included.  In Section IV, we estimate the power that 
might be radiated if such a mode were excited and the detectability of the 
resulting signal.  Discussion of the results is 
in Section V. 

The essential 
features, from KWWG, of the core-crust system, as first outlined by Alcock, 
Fahri and Olinto (1986) are: (i) A strange quark core of roughly $A/3$ each of 
u-, d-, and s-quarks with a sharp boundary on the order of a fermi where A is 
the total baryon number. The boundary is sharp because the core is bound by 
the strong force, not gravity. (ii) An electron gas extending a few hundred 
fermis beyond the core.  The
electron abundance, which is on the order of 
$10^{-4}A$, would be zero if the strange quark mass were essentially zero, as is
 true for up and down quarks, rather than the estimated $100-300$ 
MeV/${\rm c}^2$.  The electron gas beyond the core is held by, and accompanied 
by, a strong positive electric field resulting from the net positive charge on 
the sharp core; it is the gradient of a megavolt range potential. (iii) A hadron
 crust.  Non-strange matter attracted gravitationally by the core has its 
electrons repelled by the Pauli pressure of the electron gas and its ions 
repelled by the electric field.  
Neutrons suffer neither of these repulsions.  
A crust can therefore accumulate until it 
becomes energetically favorable for neutrons
to leave nuclei at the base of the crust and to ``drip" into the core. 

The mass
 of the crust is bounded by $M_C \leq 10^{-5}M_{\odot}$ for SQM star mass, 
$M_Q \simeq M_{\odot}$. KWWG note that the electrostatic potential inside the 
strange quark core is $eV(r)=\mu_e(r)$, where $\mu_e$ is the chemical potential
 for which they solve numerically along with the quark chemical potential.  
They choose specific values for the mass of the strange quark (150 MeV) and 
the MIT bag constant (50 MeV/fm), which parametrizes quark confinement in QCD.
They find, using local charge neutrality, $eV(r)$ near the surface ($r=R$) of
the core, about 18.5 MeV for zero temperature, with quadratic corrections for
finite temperature bringing $eV(r)$ near the surface down by about 0.5 MeV at
T=50 MeV.  Using global charge neutrality just at the surface, $eV(r)$, for zero
temperature, falls to 3/4 its (nearby) interior value; it falls to about half
the interior value for T=50 MeV.  
KWWG solve Poisson's equation in the gap between the
core and the hadron crust.  There are two constants of
integration.  They can be taken as the potentials at the
outer edge of the core ($r=R$) and the inner edge of the
crust ($r=R_C$).  The scale length over which the potential falls 
only depends on the first and is of the order of the few
hundred fermis.  The width of the gap is given by the value of 
$r$ at which the potential has fallen to the second.
More precisely, they show

\begin{equation}eV(r)={C \over r-R+r_0},\qquad\qquad R<r<R_C
\end{equation}
with $r_0\equiv {C\over eV(R)}$ and

\begin{equation}C=(3\pi/2)^{1/2}/e\ =\ 5\times 10^3 {\rm MeV-fm}\ =\ 8.5\times
 10^{-16} {\rm erg-cm.}
\end{equation}

\section{Vibrations}

Figure 1 shows the centers of the core and crust displaced along the polar axis
by $\xi<\Delta_G$, where $\Delta_G$ is the width of the gap.  We need to
compute the restoring force.  First, we note that for $\xi=0$, the
electrostatic repulsion and gravitational attraction balance.  The
electrostatic repulsion pressure is given by

\begin{equation}
P_{el}={Ze\eta_AC\over(\Delta_G+r_0)^2}
\end{equation}
where $\eta_A$ is the number of ions per unit area at the base of the crust and
$Ze$ is their average charge.   The gravitational attraction pressure is

\begin{equation}
P_G={GM_QM_C\over 4\pi R^4}
\end{equation}
For $\xi << (\Delta_G+r_0)$, we have

\begin{equation}
r^2=R^2_C\sin^2\phi+(R_C\cos\phi-\xi)^2;\qquad r\simeq R_C-\xi\cos\phi
\end{equation}
and so

\begin{eqnarray}F_z(\xi)&=&\left.\xi\,Ze\eta_AC\int_0^{\pi}2\pi\,\sin\phi\,
d\phi\,\frac{d^2\ }{dz^2}[r-R_Q+r_0]\ \right|_{z=0} \nonumber\\&=&-\frac{8\pi}
{3}\,\frac {\xi\,Ze\eta_AC}{(\Delta_G+r_0)^3} \\&=&-\,2/3\,\xi\,\frac{GM_QM_C}
{R^2}\,(\Delta_G+r_0)^{-1}\ \ .\nonumber
\end{eqnarray}
The result is

\begin{equation}\omega^2={2GM_Q\over 3R^2(\Delta_G+r_0)}
\end{equation}
For a strange star at zero temperature with a maximal crust, we have $\Delta_G
\sim 200$ fm and $r_0\sim 300$ fm.   From Equation (7), we can see that
$\nu_0\simeq 2.5\times 10^{11}$ Hz. and $\lambda = 1.2$ mm.   If the
temperature rises to $50$ MeV, $V(r)$ falls by $25\%$ according to KWWG, so we
then have $\nu_{50}= 2.6\times 10^{11}$ Hz and $\lambda_{50}=1.4$ mm.

We can
also ask how $\nu$ varies if $M_C$ is reduced from its maximum value when
$\Delta_G\sim 200$ fm and $\rho\simeq 4.3\times 10^{11}$ g\,cm${}^{-3}$.  The
equality of (3) and (4) gives

\begin{equation}
(\Delta_G+r_0)^2=R^4\,\beta ^2
\end{equation}
where

\begin{equation}
\beta=\left({4\pi ZeC\eta_A\over GM_QM_C}\right)^{1/2}
\end{equation}
and
\begin{equation}
\eta_A=\left({\rho\over Am_p}\right)^{2/3}
\end{equation}

In Equation (9) the quantity A is the average atomic number and $m_p$ is the 
mass of the proton. The dependence of $\rho$ on $M_C$ can be found by relating
 $\rho$ to the pressure at the base of the crust in Equation (4) by means of 
the equation of state. Using the results of Harrison and Wheeler (Harrison et 
al., 1958 and 1965) and of Baym, Pethick and Sutherland (1971) as discussed in
 Shapiro and Teukolsky(1983), we have, for $10^8<\rho<4.3\times 10^{11}=\rho_c$
, by interpolating from the numerical results,

\begin{equation}
\rho\simeq \rho_c(P/P_c)^{5/6}\qquad{\rm and}\qquad P_c =10^{29.5} {\rm
dynes/cm}^2
\end{equation}
Inserting into Equation (8), we see that $(\Delta_G+r_0)\propto M_C^{-2/9}$ and
 hence $\lambda\propto M_C^{-1/9}$.  Thus, a decrease in $M_C$ by a factor of 100
increases the wavelength by only a factor of 1.7.  

We thus find that a 
low-temperature crust of maximal mass should exhibit a signal
at about 1.2 mm; the wavelength increases with increasing temperature and
decreasing crust mass.  A 50 MeV crust with one percent of the maximal
mass would have a wavelength of about 2.4 mm.  We note that the lower
bound on the wavelength (1.2 mm)) is well above the region in which the 
atmosphere becomes opaque.

The calculation above ignores the work done against the Pauli pressure of the
electrons.  We estimate that effect using (see KWWG) for the Pauli pressure

\begin{equation}
P(r)={\mu^4\over 12\pi^2}\simeq {(4eV(r))^4\over 12\pi^2}.
\end{equation}
Taking the gradient, we find $16\pi R^2P(r)F_G^{-1}\,<\,0.05$ where $F_G\simeq
4\pi R^2P_G$.  Thus, electron Pauli pressure is a small effect.  Finally, we
note that no new modes are introduced by angular motion of the crust center
of mass about the core center of mass because of the spherical symmetry. 
However, if such motion were induced, the centrifugal term $L^2/(M_CR^3)$ would
 effectively weaken $P_G$ in Equation (4).  It would not affect $\omega^2$ 
because, as with the gravitational force, it changes on a scale of order $R$, 
not one of order $r_0$ as with the electric force.  It could, however, modulate
 the angular distribution of radiation from the system.

In summary, for strange quark star crusts of mass $M_C$ in the range
$10^{-7}M_Q < M_C < 10^{-5}M_Q$ with temperatures below about 50 MeV, the
crust-core system has a normal mode corresponding to a wavelength $\lambda$
roughly in the region $1.2$ to $2.4$ mm. 

\section{Rotational Effects}

Any real strange star will be rotating.  One expects rotational periods
ranging from milliseconds to a few seconds.  This should have two effects on
the millimeter wave signature.  The first is simply Doppler broadening--a 10
km diameter star rotating at 1 Hz would make the signal bandwidth 
$25\sin\theta$ Mhz, where $\theta$ is the angle between the rotation
axis and the observer, at
an observing frequency of 250 GHz. The bandwidth clearly scales proportional to
 the frequency.  The second effect is caused by the fact that the rotation will
cause the star to become oblate, leading to two normal modes of oscillation and
thus splitting the signal.  This is similar to the giant resonance mode in
nuclei, in which the mode is split into two modes in nuclei with spin.  In this
section, we calculate the frequency splitting.

For a non-rotating strange star, the zero-temperature equation of state is
$P={1\over 3}(\rho-4B)$, where $B$ is the bag constant.  This equation is
inserted into the Oppenheimer-Volkov (OV) equation of hydrostatic equilibrium

\begin{equation}
{d \rho\over dr}=-{(\rho+P)(m+4\pi r^3P)\over r(r-2m)}
\end{equation}
where we use units $c=G=1$ and $m$ is the mass inside the radius $r$ (and $2m$
 is its Schwarzschild radius).  The result of integrating the OV equation 
(Alcock, Farhi, Olinto, 1986) gives the structure of the star.  If it is 
rotating, there will be an additional centrifugal pressure term added to $P$ in
 the OV equation, $P\rightarrow P-{1\over 2}\rho\omega^2r^2\sin^2\theta$.  
Integrating the OV equation again with this term along the polar and equatorial
 directions gives the polar and equatorial radii, shown in Table 1, and the 
resulting eccentricity, as a function of the angular velocity of the star, 
$\Omega$.
These results were obtained in the approximation of small eccentricity. At
higher order, considerations of the mass distribution within the star,
relativistic corrections, the difference between the shape of the inner edge of
the crust and that of the core, etc. must be included.

 The strange star is now an oblate spheroid.  One will find two
normal modes, corresponding to vibrations in the polar and equatorial
directions.  We find that (again, to leading order in the eccentricity)
\begin{equation}
\Delta\omega=\left({1\over R^2_{polar}}-{1\over R^2_{equat.}}\right){2GM\over
3\beta}
\end{equation}
This frequency splitting, $\Delta\omega/\omega$, is also given in Table 1.  We
can see that for strange stars with periods of the order of seconds, the
frequency splitting is negligible relative to the Doppler broadening, whereas
for strange stars with periods of the order of milliseconds, the two are of
the same order of magnitude.  For such stars, this splitting would be a
significant indicator of a strange star origin of a narrow line, millimeter 
wave signal from a pulsar.

It should also be noted that if the pulsar is in a binary
system, the tidal distortion would also cause a splitting
of the mode, and the above equation would apply.  (If
the pulsar were rapidly rotating as well, the deformation
would be quadrupolar.)  The calculation of the tidal
force (ignoring rotation) is straightforward, and it is
easy to show that the results are identical with
$
\omega^2$ replaced by $\omega_0^2/4$, where $\omega_0$ is
the frequency of revolution.  Since the frequency of
revolution, for all realistic pulsars, is much less
than the frequency of rotation, the effects of tidal
distortion will be negligible.

\section{Radiation and Detectability}

The energy stored in the vibrational mode of Section 2 is given by
\begin{equation}
E\sim{1\over 2}M_C\xi^2\omega^2.
\end{equation}
Whereas the frequency of the mode is nearly independent of crust mass (varying
roughly as $M_C^{1/9}$), the energy stored goes roughly as the $11/9$ power.
For $M_C\sim 10^{-5}M_{\odot}$ and $\xi\sim 200$ fm, E is of the order of
$10^{31}$ ergs.  We estimate the radiation rate in a simple model.  When the
crust center of mass is displaced downward, relative to the core center of mass
by $\xi$, as shown in Figure 1, the electrostatic potential at the ``top" rises
by $-\xi{dV\over dx}>0$ and the potential at the ``bottom" falls by the same
amount.  However, the crust is an equipotential, made from a material of very
high conductivity, although likely not a superconductor, so charge must flow to
cancel this change.  Consider the ``flat star approximation" in which the
crust  consists of two parallel planes, each of radius $R$ and with separation
$R$.  To maintain the equipotential, a sheet of charge density $\sigma$ must
flow with current $I=\omega Q$,
\begin{equation}
\Delta V=4\pi\sigma R=2\xi{dV\over dz},
\end{equation}
and
\begin{equation}
Q=\pi R^2\sigma=R{\Delta V\over 4},
\end{equation}
giving radiation power on the order of \begin{equation}
P={dE\over dt}={I^2\over 2c}\simeq (\omega\xi RV')^2
\end{equation}

This expression will only be valid for temperature not too much smaller than 
the 5 MeV Fermi momentum of the electrons in the crust.  For very small T, we 
would expect the radiation rate to fall like $(T/p_F)^3$ as Pauli blocking makes
 electrons (and holes) unable to radiate at frequency $\nu$.  For 
$\xi V' \sim 10$ MeV/e, Equation (18) gives 
$P\sim (10\,eV/e)^2(R\omega)^2/c\sim 10^{34}$ erg/s, which is a very large 
signal.  This rate is reduced to the extent that radiation from electrons not 
at the surface will either not occur or will be absorbed and simply heat the 
crust.  One rough estimate of this reduction would be to assume that the charge
 is spread evenly throughout the crust and hence reduce the intensity of the 
radiation by $\lambda/\Delta_C\sim10^{-5}$, where $\Delta_C\sim 100m$ is the 
crust thickness. However the crust conductivity is so high, 
$\sigma\sim10^{25}s^{-1}$, (Pethick and Sahrling, 1995) that the skin depth 
(Jackson,1975) is far less than the wave length.  For a conservative estimate 
of this effect we take the geometric mean between $\lambda$ and 
$\lambda /\Delta_C$ giving $P_{rad}\sim10^{-3}P.$ Thus, a rough power estimate
 might be $10^{31}$ erg/s, for maximal excitation of the mode and for a crust
 of maximum thickness.  The power, but not the decay time, would scale with the
 energy in the mode. Roughly, the decay time would scale with the thickness of 
the crust and the power radiated would scale inversely.

 At a distance $D$ in 
kpc from the pulsar, the flux density of radiation emitted at the rate of $P$ 
Watts spread over a frequency $f$ (Hz) is
\begin{equation}S=8.3\times 10^{-15}\frac{P}{fD^2}\ {\rm Jy}
\end{equation}
where 1 Jy is $10^{26}$ W/m${}^2$/Hz.  A 10 meter sub-millimeter telescope 
operating at a frequency of $250$ GHz
 in good conditions near zenith has an rms noise
\begin{equation}\Delta S={7000\over (ft)^{1/2}}\rm{Jy}
\end{equation}
where the integration time, $t$, is in seconds.  The signal to noise ratio for
 a continuous signal is
\begin{equation}
{S\over \Delta S}=1.2\times 10^{-18}\ P\left({t\over f}\right)^{1/2}
{1\over D^2}
\end{equation}
For a pulsar of period 1 second and a consequential Doppler broadening of 
$\sim 25$ MHz (at an observing frequency of 250 GHz) emitting a $10^{24}$ W 
signal, a 15 second integration time yields a signal-to-noise ratio 
$\sim 10^3 D^{-2}$.  A detectable signal ($5\sigma$) could be achieved for 
pulsars as far away as 15 kpc (there are more than 100 with periods in excess 
of 1 second).  For a pulsar as close as 1 kpc (there are about ten with periods
 longer than 1 second) a $10^{29}$ erg s${}^{-1}$ signal could be detected.  
The strength of the signal is not the problem for detection; the problem is not
 knowing the frequency.  Receivers in the 250 GHz range have instantaneous
 bandwidths of a few GHz which could be searched in a few minutes with a 1 GHz 
spectrometer.  Since retuning the receiver to the next band would take perhaps 
a quarter hour, most of the time for the search would be spent in retuning the 
receivers, not in the observing.

\ \section{Discussion}
One of the most important questions in the study of compact objects is the
nature of pulsars, including whether pulsars consist of neutron matter or 
strange quark matter.  In this paper, we have identified an observable radio 
signal that would be characteristic of the latter possibility.  A spectral line
originating from the pulsar could easily be distinguished from one
originating in the interstellar medium because the pulsar line will
be detected only when the telescope is pointed at the pulsar.
 We have, however, left important questions 
unanswered. How is the vibrational mode excited?  A transient signal, such as 
that due to a starquake or cometary impact, would die off quickly, on the order
 of milliseconds, and would thus be unobservable.  Does a mechanism for 
continuous excitation exist?  The power of a detectable signal is only on the 
order of $10^{-6}$ of pulsar energy losses.  Thus it would seem reasonable, in
 view of the many consequences, to make observations -- independent of 
theoretical considerations, in order to see if such a small fraction of the 
energy loss does, in fact, go into this mode. We may speculate, however, that 
one direction from which such a mechanism could come might be that of the 
interaction between superfluid vortices and (type 2) superconducting magnetic 
flux tubes. (We know that the magnetic field for a strange quark star must pass
 through the core because the crust is so thin.) It is the outward migration of
 the superfluid vortices that is responsible for pulsar spin-down, while 
Ruderman (1996) has conjectured that the latter could be effectively pinned by 
the interaction with the non-superconducting electron fluid.  Some effective 
resonant coupling between those two systems, on the one hand, and the 
vibrational mode, on the other, could result in continuous excitation.  For 
example, the coupling could be associated with discontinuous passage of a 
vortex through a pinned flux tube. 

Even if the mode is excited, we would 
expect some fraction of the energy stored in the mode to go into heating the 
crust, and some fraction to go into radiation.  Our crude estimate that 
$P_{rad}/P\sim 10^{-3}$ of the energy goes into radiation needs to be improved.
  Another question concerns the scale over which the coherence implicit in our 
calculations is maintained. If such coherence is only maintained over some 
fraction of the size of the star, then the power radiated would be reduced by 
that fraction.  

The signal predicted in this paper is a speculative one; however the dearth of
distinctive signatures for strange quark stars, in our view, makes the search 
for such a signal worthwhile.\\[1.cm]

It is a pleasure for VLT to acknowledge a very helpful and enjoyable series of
 conversations on both neutron stars and strange quark stars with Mal Ruderman;
VLT and DCR are grateful to O. W. Greenberg for the hospitality of the Physics 
Department of the University of Maryland during the 1995-96 academic year when
 part of this work was done.  VLT is also grateful for  useful conversations with 
Virginia Trimble and Duane Dicus.

\newpage\centerline{\bf References}
\parindent=0pt
\vskip .5cm
Alcock, C., Farhi, E., \& Olinto, A. 1986, ApJ, 310, 261
\vskip .5cm
Baym, G, Pethick, C., \& Sutherland, P. 1971, ApJ, 170, 299
\vskip .5cm
Benvenuto, O.G., \& Vucetich, H. 1991, Nucl. Phys. (Proc. Suppl.) 24B, 160
\vskip .5cm
Bodner, A.R. 1971, Phys. Rev., D4, 160
\vskip .5cm
Cottingham, W.N., Kalafatis, D., \& Vinh Mau, R. 1994, Phys. Rev. Lett, 73,1328
\vskip .5cm
Pethick, C.P., \& Sahrling, M. 1995, ApJ, 453, L29
\vskip .5cm

de Rujula, A., and Glashow, S.L.,1984, 2nd Int. Symp. on Resonace Ionization
Spectroscopy and Applications, Knoxville TN, April 1984
\vskip .5cm
Farhi, E., and Jaffe, R.L.,
1984, Phys. Rev. D30, 2379
\vskip .5cm
Friedman, B., and McLerran, L., 1978, Phys. Rev. D17,
1109 
\vskip .5cm
Frieman, J.A., and Olinto, A., 1989, Nature 341, 633
\vskip .5cm
Glendenning, N.K., 1989, Phys. Rev. Lett. 63, 2629
\vskip .5cm
Harrison, B.K., Thorne, K.S., Wakano, M, and Wheeler, J.A., 1965, "Gravitational
Theory and Gravitational Collapse",  Univ. of Chicago Press
\vskip .5cm
Kettner, C., Weber, F., Weigel, M.K., and Glendenning, N.K., 1995, Phys. Rev.
D51, 1440
\vskip .5cm
Madsen J., and Haensel, P., 1991, Nucl. Phys. (Proc. Suppl.) 24B, 1
\vskip .5cm
Ruderman, M., 1996, private communication
\vskip .5cm
Shapiro, S.L., and Teukolsky, S.A., 1983, "Black Holes, White Dwarfs, and
Neutron Stars:  the Physics of Compact Objects", Wiley, NY
\vskip .5cm
Usov, V.V., 1997, ApJ 481, 1107
\vskip .5cm
Witten, E., 1984, Phys. Rev. D30, 272
\vskip .5cm
\newpage
\centerline{\bf TABLE 1}
\vskip 3.0cm

\begin{tabular}{|c|c|c|c|}
\hline
&&&\\
Rotational&Doppler&eccentricity&Mode\\
Frequency(Hz)&Broadening(GHz)&&Splitting(GHz)\\
&&&\\
\hline
&&&\\
$10$&$.25$&$6\times 10^{-6}$&$.0015$\\
&&&\\
\hline
&&&\\
$100$&$2.5$&$5.5\times 10^{-4}$&$0.13$\\
&&&\\
\hline
&&&\\
$1000$&$25$&$.058$&$15$\\
\hline
\end{tabular}
\vskip 3.0cm
{\bf Table 1}  For strange stars rotating with angular velocities
from 1-1000 Hz, we give the expected Doppler broadening of the
250 GHz signal, the eccentricity of the star and the resulting splitting
of the signal.  We give results for a strange star 
central density of $5.5B$, where $B$ is the bag constant;
the results of the mode splitting will vary by roughly a factor of
two for the expected range of central densities.
  The Doppler broadening results
assume the observer is in the equatorial plane; if this isn't the case,
the results should be multiplied by $\sin\theta$.
\newpage
\centerline{\bf Figure Caption}
\vskip 3.0cm

{\bf Figure 1}.  The figure shows the centers of the core and crust
being displaced.  The resulting oscillation has a frequency of 
approximately 250 GHz, leading to the millimeter wave signal discussed
in the paper.

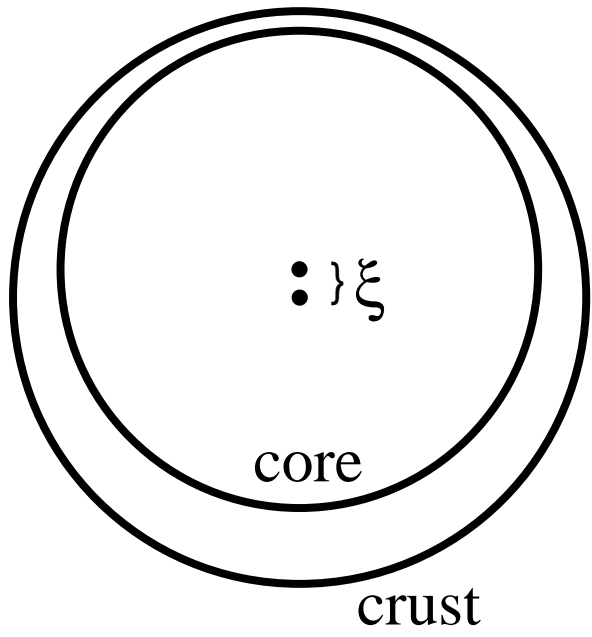
\end{document}